\documentclass[11pt,onecolumn,draftclsnofoot]{IEEEtran}
\usepackage{amsfonts}
\IEEEoverridecommandlockouts

\ifCLASSINFOpdf
\else
\fi
\usepackage{psfig}
\usepackage{epsf}
\usepackage{bm} 
\usepackage{slashbox}
\usepackage{color}
\usepackage{booktabs}
\usepackage{pifont}
\usepackage{times,amsmath,color,amssymb,graphicx,geometry, epsfig,psfrag,subfigure,algorithm}
\usepackage{enumerate}

\begin{document}
\title{Energy-Efficient NOMA Enabled Heterogeneous Cloud Radio Access Networks}
\author{\IEEEauthorblockN{Fuhui Zhou, \emph{Member, IEEE}, Yongpeng Wu, \emph{Senior Member, IEEE}, Rose Qingyang Hu, \emph{Senior Member, IEEE}, Yuhao Wang, \emph{Senior Member, IEEE}, Kai-Kit Wong, \emph{Fellow, IEEE}}
\thanks{Fuhui Zhou is with the School of Information Engineering and with Postdoctoral Research Station of Environmental Science and Engineering, Nanchang University, P. R. China, 330031, and also with the Department of Electrical and Computer Engineering at Utah State University, Logan, UT 84322 USA (e-mail: zhoufuhui@ncu.edu.cn).

Yongpeng Wu is with the Department of Electronic Engineering, Shanghai Jiao Tong University, P. R. China, 200240 (e-mail: yongpeng.wu2016@gmail.com).

Rose Qingyang Hu is with Electrical and Computer Engineering Department, Utah State University, Logan, UT 84322 USA (e-mail: rose.hu@usu.edu).

Yuhao Wang is with the School of Information Engineering, Nanchang University, P. R. China, 330031 (e-mail: wangyuhao@ncu.edu.cn).

Kat-Kit Wong is with the Department of Electronic and Electrical Engineering, University College London, WC1E 7JE, United Kingdom (e-mail: kai-kit.wong@ucl.ac.uk).
}
\thanks{The research was supported by the Natural Science Foundation of China (61701214, 61661028, and 61561034), The Young Natural Science Foundation of Jiangxi Province (20171BAB212002),  the China Postdoctoral Science Foundation (2017M610400), the Key Programs for Young Natural Science Foundation of Jiangxi Province (20152ACB21008) and Young Scientist of Jiangxi Province(20142BCB23001), the Natural Science Foundation of Jiangxi Province(2015BAB207001).}}
\maketitle
\begin{abstract}
Heterogeneous cloud radio access networks (H-CRANs) are envisioned to be promising in the fifth generation (5G) wireless networks. H-CRANs enable users to enjoy diverse services with high energy efficiency, high spectral efficiency, and low-cost operation, which are achieved by using cloud computing and virtualization techniques. However, H-CRANs face many technical challenges due to massive user connectivity, increasingly severe spectrum scarcity and energy-constrained devices. These challenges may significantly decrease the quality of service of users if not properly tackled. Non-orthogonal multiple access (NOMA) schemes exploit non-orthogonal resources to provide services for multiple users and are receiving increasing attention for their potential of improving spectral and energy efficiency in 5G networks. In this article a framework for energy-efficient NOMA H-CRANs is presented. The enabling technologies for NOMA H-CRANs are surveyed. Challenges to implement these technologies and open issues are discussed. This article also presents the performance evaluation on energy efficiency of H-CRANs with NOMA.
\end{abstract}

\IEEEpeerreviewmaketitle
\section{Introduction}
\IEEEPARstart{T}{HE} explosive increase of the number of smart devices, such as smart phones and smart tablets, as well as emerging wideband and high-rate services, such as augmented reality (AR) and virtual reality (VR), and the massive number of devices constructing the Internet of Things (IoT), enable an urgency for designing energy-efficient communication systems in order to achieve environmentally friendly, greenly economic, and sustainable operations. Compared with the fourth generation (4G) systems, the fifth generation (5G) systems are required to achieve 1000 times higher system capacity, 10 times higher spectral efficiency (SE), at least 100 times higher energy efficiency (EE), 1 ms latency, and 100 times higher connectivity density \cite{J. G. Andrews}. As a promising new technology and architecture, heterogeneous cloud radio access networks (H-CRANs) have drawn significant attention in both industry and academia. H-CRANs aim to achieve high flexibility, tremendous capacity, high EE, wide coverage, and cost-effective operation \cite{R. Q. Hu}-\cite{G. Wang} mainly by incorporating powerful cloud computing and virtualization techniques into heterogeneous networks (HetNets).

Different from HetNets, in H-CRANs, base stations from different tiers (e.g., macro cells, micro cells, pico cells and femto cells) are decoupled into baseband units (BBUs) and remote radio heads (RRHs). All BBUs construct a BBU pool in a cloud center, and RRHs are deployed close to user equipments (UEs). The BBU pool efficiently performs baseband signal processing (e.g., modulation, coding, radio resource allocation, media access control, etc.) through cloud computing and virtualization techniques, while the RRHs remotely conduct radio transmission/reception processing and convert the radio signals to digital base band signals and vice versa. H-CRANs can greatly increase the flexibility of the network architecture, improve the system SE, and significantly reduce energy consumption and operational expenditures. Moreover, the quality of service (QoS) of users can be remarkably improved due to the reduced distance between  RRHs and UEs as well as  RRH association from different tiers.

It is envisioned that numerous multiple access technologies will be employed in future H-CRANs in order to mitigate inter and intra cell interference and to improve SE and EE. As a new multiple access technique, non-orthogonal multiple access (NOMA) has been identified as a promising candidate for significantly improving SE and EE of 5G mobile communication networks \cite{Z. Ding}. Unlike the conventional orthogonal multiple access (OMA) schemes, NOMA schemes (such as power-domain NOMA, code-domain NOMA and sparse code multiple access) provide services for multiple users by using the non-orthogonal resources. For example, the power-domain NOMA exploits different power levels to provide services for multiple users at the same frequency band and the same time. At receivers, successive interference cancellation (SIC) is applied to decrease the mutual interference caused by using the non-orthogonal resources. The non-orthogonality enables NOMA techniques to have advantages in supporting high SE and EE, massive connectivity, and low transmission latency at the cost of NOMA interference and the complexity of users' receivers \cite{L. Dai}-\cite{Y. Zhang}.

While H-CRANs have been extensively studied, there has not been much relevant work on NOMA H-CRANs. In \cite{R. Q. Hu}, an energy-efficient H-CRAN framework was established. It was shown that both EE and SE can be significantly improved by using H-CRANs. The authors in \cite{M. Peng} discussed five key techniques based on cloud computing for H-CRANs. In \cite{H. Dahrouj}, challenges for three promising resource allocation schemes for H-CRANs were investigated. Resource sharing in H-CRANs at three different levels (spectrum, infrastructure, and network) was analyzed in \cite{M. A. Marotta}. None of these existing works on H-CRANs investigated multiple access techniques, which actually are of great importance in H-CRANs for interference mitigation, EE and SE improvement, and low-cost operation. Thus, different from previous works, this article focuses on the enabling technologies and challenges of employing NOMA in H-CRANs.

\begin{figure}[!t]
\centering
\includegraphics[width=5.0 in]{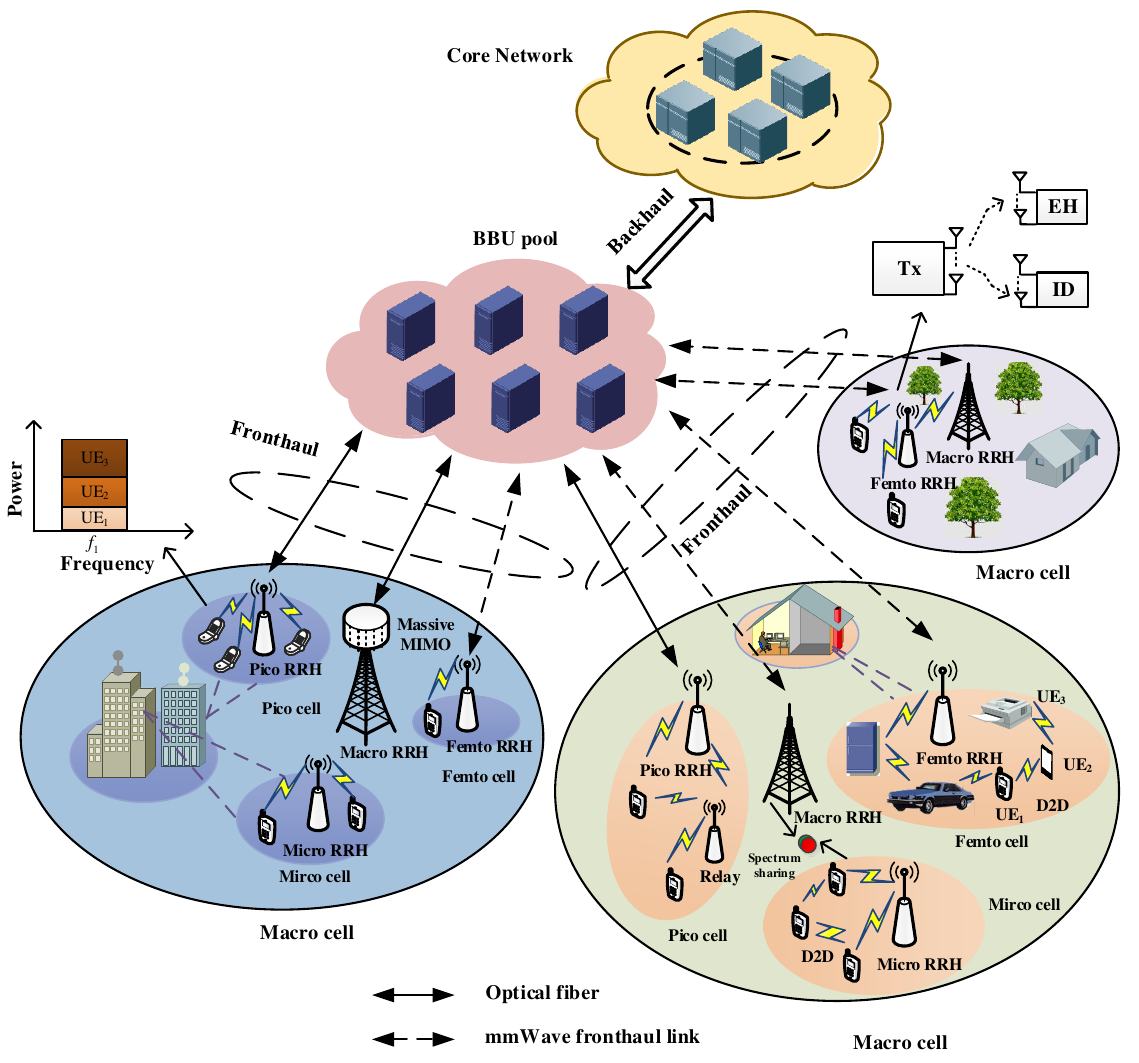}
\caption{System Architecture of a NOMA H-CRAN.} \label{fig.1}
\end{figure}
The rest of the article is organized as follows. In Section II, the H-CRAN architecture and the implementation of NOMA are discussed. In Section III, the promising enabling techniques, particularly related to applying NOMA to H-CRANs are elaborated. These technologies include massive multiple-input multiple-output (MIMO), cognitive radio (CR), millimeter-wave (mmW) communications, wireless charging, cooperative transmission, and device-to-device (D2D) communications. Challenges and open issues for these technologies are presented in Section IV. The article concludes with Section V.
\section{NOMA H-CRAN Architecture}
The system architecture of a NOMA H-CRAN is shown in Fig. 1. It consists of macro-cells, micro-cells, pico-cells, and femto-cells. Macro-cells or micro-cells with high power antennas implemented  as remote radio heads (RRHs) provide large coverage areas but relatively low data rates (e.g., in urban, suburban and rural areas), while pico-cells or femto-cells with low power RRHs  are deployed in small coverage areas requiring high data rate transmission such as airports, stadiums and homes. Unlike the conventional HetNets,  radio resource allocation and media access control in H-CRANs are performed in the BBU pool. The links between RRHs and the BBU pool can be wired (e.g., optical fiber) or wireless (e.g., millimeter-wave). The best choice depends on the operational cost and the wireless channel condition since the cost of deploying optical fiber links can be high and the impact of the wireless channel conditions on the achievable capacity is critical. It is expected that both wired and wireless links will be exploited in future H-CRANs.

In order to cater to diverse service requirements, different applications, and cells with different scales, and also to improve the operational efficiency, OMA and NOMA can coexist in and can be used both in the uplink and downlink of H-CRANs. Specifically, OMA is appropriate for femto-cells, where the number of UEs is normally low and real-time services with high data rates are required. Furthermore, in the small area like femto-cell, channel diversity can be limited, making OMA a more suitable choice. On the other hand, NOMA is a desirable candidate for cells that require a high connectivity density and frequent small data rate transmissions, e.g., in large-scale shopping centers or some Internet of Things (IoT) type applications. The combination of OMA and NOMA is appropriate in scenarios where both massive UEs connectivity and high rates are required. For example, the combination of orthogonal frequency division multiple access (OFDMA) and NOMA is well suitable for multiple neighboring femto-cells, where different subbands are allocated to different femto-cells and each subband exploits NOMA to provide services for one femto-cell. Moreover, different NOMA schemes and OMA schemes have their own appropriate application scenarios, which depend on the tradeoff between the achievable performance and implementation complexity. As an example, power-domain NOMA  can be used if the channel conditions from different RRHs to UEs have sufficient diversity. The authors in \cite{L. Dai} have comprehensively discussed the applied situations of different NOMA schemes.

The physical layer framework of a downlink NOMA H-CRAN is presented in Fig. 2(a). The BBU pool performs adaptive modulation and coding (AMC) for the $N$ data sources that are used to provide services for $N$ UEs. Then, the BBU pool allocates non-orthogonal resources to different data streams. This operation is based on the system design requirement and the channel state information (CSI) feedback. When each UE receives the transmitted NOMA signal, each UE performs multi-user detection based on a decoding order. The decoded signal can be removed from the superimposed received signals. This process continues until all the individual signals in the superimposed NOMA signals are decoded under the decoding order.
\section{Enabling Technologies in NOMA H-CRANs}
In this section, we discuss in detail enabling techniques for NOMA H-CRANs and  performance studies are presented to give more insights on different  technologies. Then, in the following section, challenges and open issues for the implementation of these technologies in NOMA H-CRANs are discussed.
\subsection{Massive MIMO with NOMA }
\begin{figure}[htb]
\centering
\includegraphics[width=3.2 in,height=3.3 in]{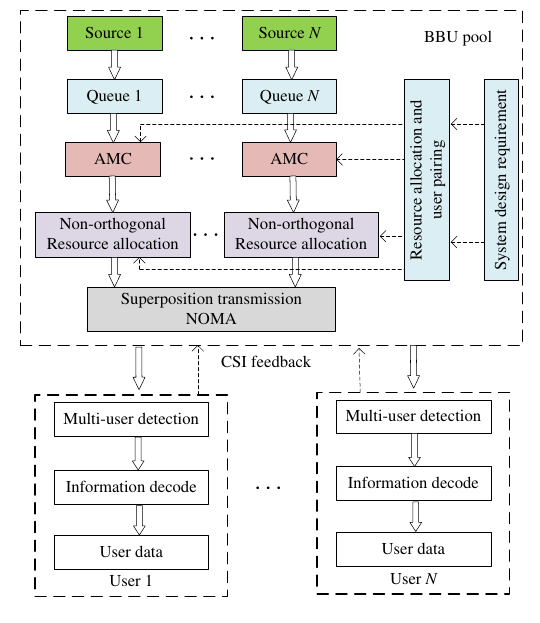}
\includegraphics[width=3.2 in,height=3.3 in]{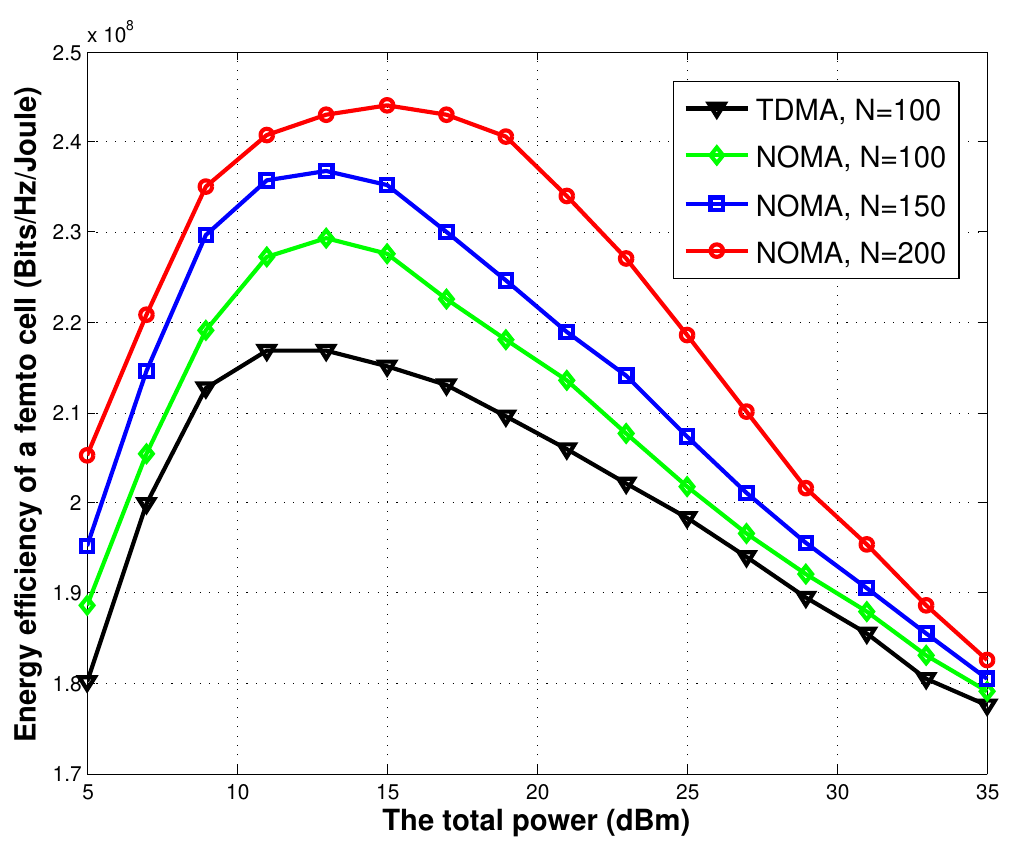}
\put(-310,-10){\footnotesize{(a)}}
\put(-100,-10){\footnotesize{(b)}}
\caption{ (a) The physical-layer framework of a downlink NOMA  H-CRAN; (b) The total downlink EE of a femto cell with five downlink UEs versus the total transmit power of the RRHs for different numbers of antennas, $N=100, 150, 200$.} \label{fig1}
\end{figure}
Applying NOMA in H-CRANs causes extra interference among UEs due to the usage of non-orthogonal resources. Although the interference can be partially mitigated by conducting baseband signal processing in the BBU pool using cloud computing techniques, the required communication overhead between diverse RRHs and the BBU pool can be extremely high. One promising scheme is to use massive MIMO, which equips RRHs with a higher number of antennas (e.g., 100 or more). Massive MIMO with NOMA can significantly improve EE and SE of NOMA H-CRANs while requiring only simple linear signal processing approaches \cite{L. Lu}. Moreover, unlike in the conventional NOMA HetNets, multi-user detection (e.g., SIC for power domain NOMA, a message passing algorithm for code domain NOMA) can be moved into the BBU pool, which greatly simplifies the structure of RRHs and makes it possible for RRHs to be cost-efficiently deployed in a large-scale area. Furthermore,  massive MIMO with NOMA can provide a large number of degrees of freedom and suppress inter-cells interference by performing precoding of UEs' signals in different cells.

In order to evaluate the superiority of massive MIMO with NOMA, Fig. 2(b) is given to compare EE achieved by using the power-domain NOMA scheme with that obtained by using the time-division multiple access (TDMA) scheme in a downlink femto-cell. EE is defined as the ratio of the total rate to the total power consumed by RRHs. The number of UEs is $5$ and the number of downlink RRHs is $3$. The constant power consumption of RRHs is 10 dBm. The number of antennas of each RRH is set to $100$, $150$ and $200$. The distance between the RRHs is $100$ m, and UEs are uniformly distributed. From Fig. 2(b) one can see that the EE achieved by usig NOMA is larger than that obtained by using TDMA. This indicates that NOMA is superior to OMA in terms of EE. Moreover, as shown in Fig. 2(b), the maximum EE increases with the number of antennas of RRHs as the diversity gain increases  when the number of RRH antennas  increases.
\subsection{Cognitive Radio with NOMA}
CR is envisaged to allievate the increasingly severer spectrum scarcity problem  due to reasons including  fixed spectrum allocation strategy, the ubiquitous wireless devices access, and the ever increasing capacity requirements. CR enables unlicensed UEs to use the spectrum resources of licensed UEs in an opportunistic mode, spectrum-sharing mode, or in a sensing-sharing mode. SE can be greatly enhanced with CR. Several standards have been proposed for CR, including IEEE 802.22, IEEE 802.11 TGaf, IEEE 802.16h, and IEEE 802.19. Recently, the exploitation of NOMA in CR has been considered for 3GPP-LTE. It was shown that both SE and EE can be improved by using NOMA in CR networks compared with these achieved by using the conventional OMA \cite{Y. Liu}. Moreover, the combination of NOMA and CR is facilitated in H-CRANs, where spectrum sensing, spectrum assignment, power allocation, and interference control can be efficiently performed in the BBU pool with powerful cloud computing techniques. An additional advantage of applying NOMA with CR in H-CRANs is that multiple unlicensed UEs can be served simultaneously, which is unrealistic for OMA unless multiple frequency bands are available.
\begin{figure}[!t]
\centering
\includegraphics[width=6 in]{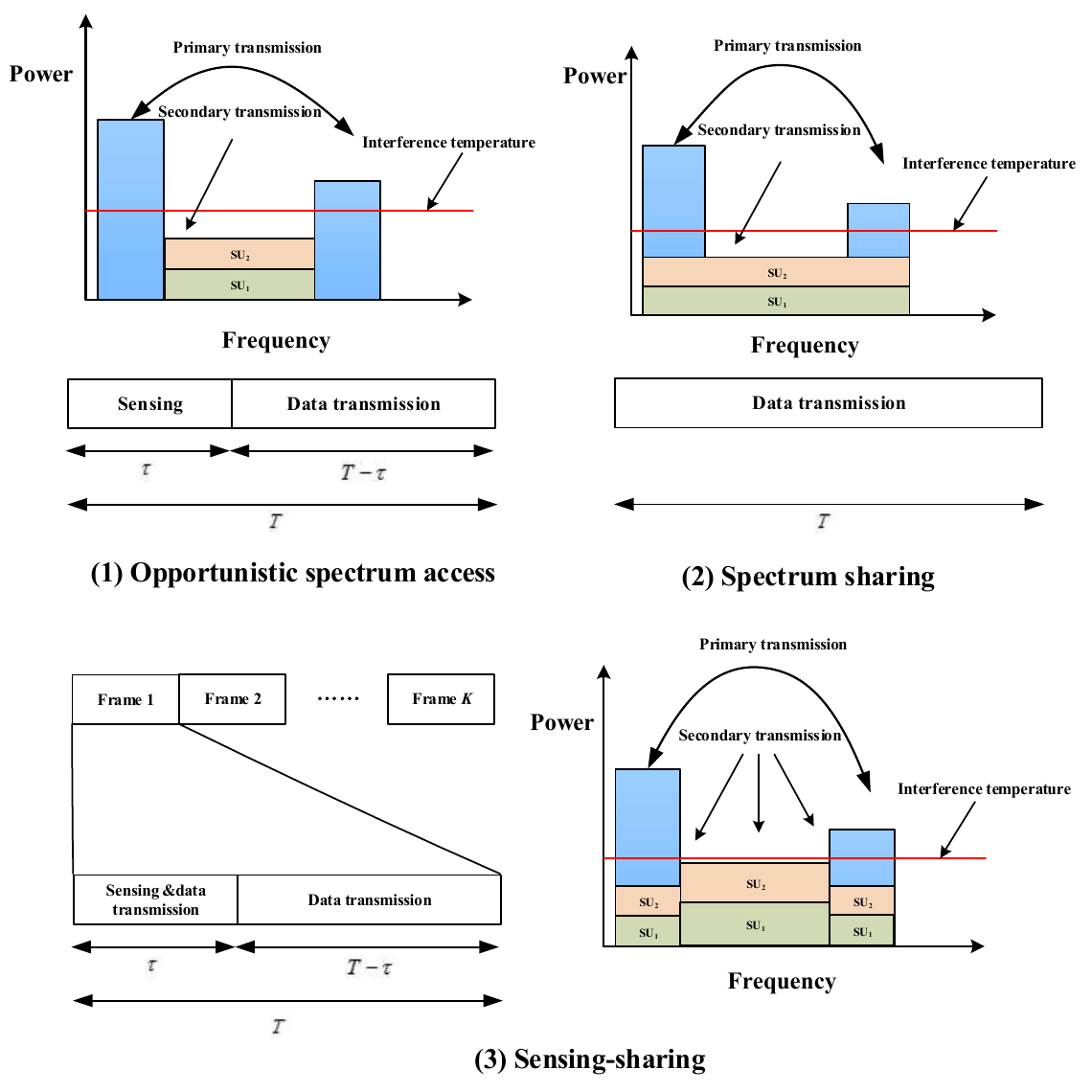}
\caption{The frame structure and operation paradigm of the combination of power-domain NOMA and CR in H-CRANs. } \label{fig.1}
\end{figure}

There are three operation paradigms for realizing the combination of NOMA with CR in H-CRANs. Fig. 3 shows the frame structure and the corresponding operation paradigm of the combination of power-domain NOMA and CR. Under the opportunistic spectrum access, the frame structure consists of a sensing slot and a data transmission slot. Multiple unlicensed UEs can simultaneously access the frequency bands owned by the licensed UEs using NOMA techniques only when the licensed UEs are detected to be inactive. In this case, high performance spectrum sensing algorithms are required in the BBU pool, and the proper choice of the sensing duration is critical  to achieve a good tradeoff between the quality of service (QoS) of the licensed UEs and the total throughput of the unlicensed UEs. Under the spectrum-sharing mode, since the unlicensed UEs can coexist with the licensed UEs as long as the interference caused to the licensed UEs is tolerable, the frame structure only consists of the data transmission slot. In this mode, an appropriate power allocation strategy of the unlicensed UEs is crucial so that the QoS of the licensed UEs can be protected and the unlicensed UEs can achieve a good performance. Finally, under the sensing-sharing mode, which is a hybrid mode of the opportunistic access and spectrum sharing, the frame structure  also consists of a sensing slot and a data transmission slot. In this mode, the joint design of spectrum sensing algorithms, the sensing duration, and the power allocation strategy is vital to obtain a high SE.

The optimal selection of the CR operation mode in NOMA H-CRANs depends on the tradeoff between the achievable performance and the implementation complexity of the BBU pool. In particular, the sensing-sharing mode can be selected if a high implementation complexity can be afforded in the BBU pool. The reason is that it can provide a good performance due to the flexible power allocation strategy based on the spectrum sensing result, that is, a high transmit power level of dowlink RRHs or uplink UEs can be used when the licensed UEs are detected to be inactive, whereas a low transmit power level is chosen when the licensed UEs are detected to be active. If  a low implementation complexity is preferred,  the spectrum sharing mode is selected due to its facilitated implementation. The opportunistic access mode can be selected in the case that the interference  to the licensed UEs can be strictly controlled (e.g., IEEE 802.22 for the TV bands).

Fig. 4(a) shows the EE of a NOMA cognitive micro-cell versus the maximum transmission power of the unlicensed RRH. The spectrum sharing mode is applied and the TDMA scheme is used as the benchmark for comparison. The simulation results are obtained under the EE/SE maximization objectives with the constraints that the minimum rate of unlicensed UEs is guaranteed and  that the maximum tolerable interference power of licensed UEs is satisfied. The number of the unlicensed RRH's antennas is 5 and the number of the licensed RRH's antennas is 3. The minimum capacity of unlicensed UEs is set as 3 Bits/s/Hz. The interference power is set as 0.1 W or 0.2 W. The variance of noise is 1. The transmission power of the licensed RRH is 1 W. The number of the unlicensed UEs is 3 and the number of the licensed UEs is 2. All the channels are modeled with  Rayleigh flat fading. The constant power consumption is 0.1 W. It is seen from Fig. 4(a) that NOMA can provide a higher EE gain than TDMA. The study clearly shows that there is  a performance tradeoff between EE and SE.
\begin{figure}[htb]
\centering
\includegraphics[width=3.2 in,height=3.0 in]{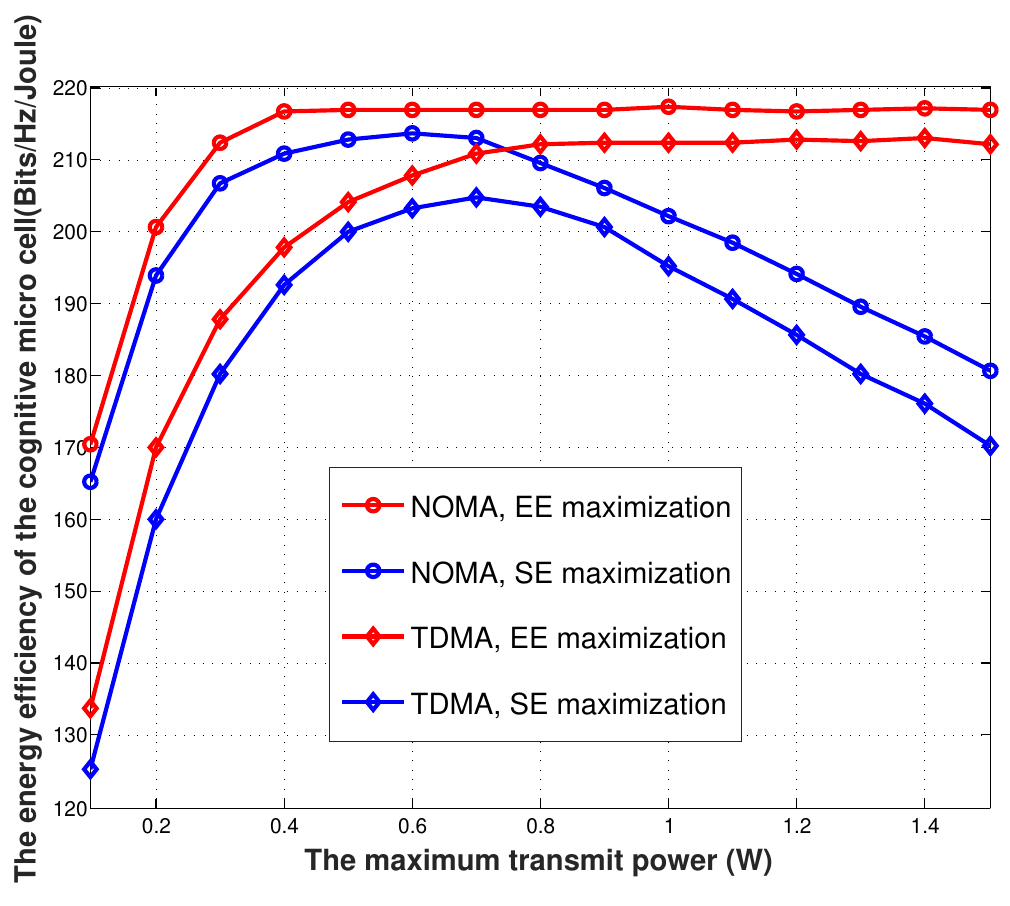}
\includegraphics[width=3.2 in,height=3.1 in]{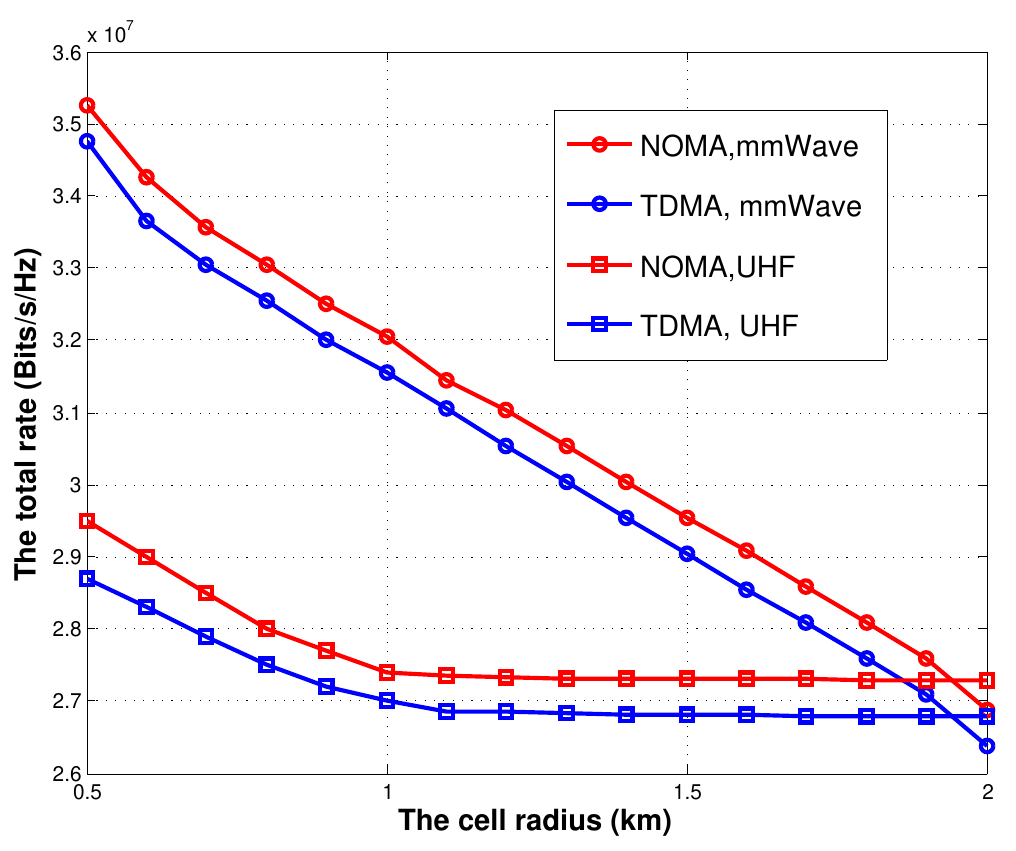}
\put(-310,-10){\footnotesize{(a)}}
\put(-100,-10){\footnotesize{(b)}}
\caption{ (a) EE of the cognitive micro-cell versus the maximum transmission power of the secondary RRH; (b) The average total rate versus the cell radius achieved by using NOMA and TDMA under the mmWave channel and the UHF channel.} \label{fig1}
\end{figure}
\subsection{Millimeter-Wave Communications with NOMA}
The availability of 28, 38, 60 and 73 GHz bands for communications has attracted significant attention to mmWave communications, which normally operate on mmWave bands between 30 and 300 GHz. mmWave communications are considered as an important technique for 5G due to its great  potentials for providing ultra-wide bands services, enabling the deployment of large numbers of miniaturized antennas ($\geq$ 32) in a small dimension, and allowing massive connectivity of different devices with diverse service requirements \cite{S. Rangan}. It can be foreseen that the combination of mmWave communications with NOMA in H-CRANs can further improve SE and accommodate  more device connectivity. This combination is deemed appropriate for the urban outdoor environment due to the high UEs density, small cell radii (about 100-200 m), and lower mobility. MmWave is also a desirable technology  for the fronthaul links between RRHs and the BBU pool,  where  wide-band services and tremendous capacities are needed.

Several standardization activities have been devoted to implementing the combination of mmWave with NOMA, e.g. Rel-13 3GPP. Since the free-space propagation loss depends on the operating frequency, different mmWave frequency bands are appropriate for different services. For example, 57-64 GHz is promoted to provide orders of magnitude gigabit data rates in wireless local area networks. Additionally, spectrum bands up to 252 GHz is potentially suitable for mobile broadband services. In order to give valuable insights on the effect of the free-space propagation on the performance of mmWave with NOMA, Fig. 4(b)  shows the average total rate versus the the cell radius achieved by using NOMA and TDMA under the mmWave channel and  ultra high frequency (UHF) channel. The path loss model for mmWave channel is ${L_{mm}}\left( r \right) = \rho  + 10\beta {\log _{10}}\left( r \right) + {\chi _{mm}}$, where $\rho  = 32.4 + 20{\log _{10}}\left( {{f_c}} \right)$; $r$ is the distance between UE and  RRH; $f_c$ is the carrier frequency and ${\chi _{mm}}$ is the zero mean log normal random variable for the mmWave link. $\beta$ denotes the path loss exponents \cite{S. Rangan}. The path loss model for the UHF channel is ${L_{UHF}}\left( r \right) = 20{\log _{10}}\left( {\frac{{4\pi }}{{{\lambda _c}}}} \right) + 10\beta {\log _{10}}\left( r \right) + {\chi _{UHF}}$, where $\chi _{UHF}$ is the shadow fading.

The simulation results are obtained in a mmWave NOMA uplink where two UEs transmit information to a RRH.  TDMA over the UHF channel is used for comparison. $\beta$ is 3.3 for the mmWave link while $\beta$ is 2.5 for the UHF link. The UEs' transmission power is 24 dBm. It is assumed that UEs are randomly distributed. For the UHF link, the carrier frequency is 3 GHz and the bandwidth is 10 MHz while the carrier frequency of the mmWave link is 94 GHz and the bandwidth is 2GHz. It is seen from Fig. 4(b) that NOMA is superior to TDMA with respect to SE for both the mmWave link  and the UHF link. It is also seen that the performance of mmWave link is greatly influenced by UE locations.  The reason is that the path loss increases with the distance between UE and RRH. This indicates that mmWave techniques are more appropriate for short distance communications.

\subsection{Wireless Charging and NOMA}
In H-CRANs, a major limitation of the network performance can be battery driven energy-constrained devices such as  energy-limited wireless sensor, mobile phone, electrical vehicles. Wireless charging techniques that enable energy-constrained devices to replenish energy from the surrounding electromagnetic radiations are deemed promising solutions to conquer this limitation \cite{E. Boshkovska}. An advantage of exploring wireless charging techniques in H-CRANs is the high power transfer efficiency due to the short distances among RRHs and UEs. Moreover, when wireless charging and NOMA techniques are applied in H-CRANs, NOMA UEs (say group 1) that are close to RRHs can also be energy sources for NOMA UEs (say group 2) far away from RRHs but close to group 1 UEs.
\begin{figure}[htb]
\centering
\includegraphics[width=3.2 in,height=3.0 in]{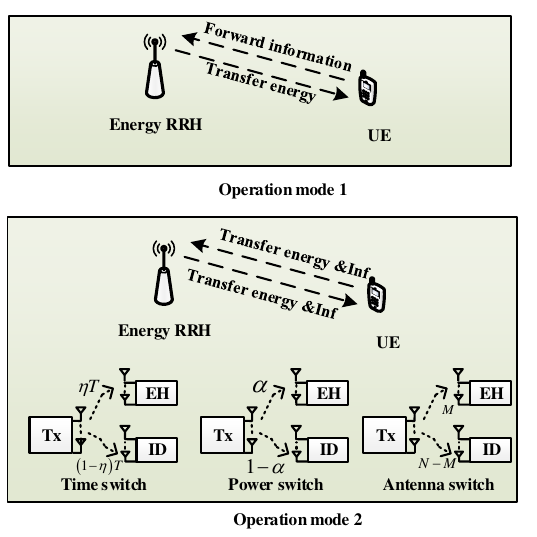}
\includegraphics[width=3.2 in,height=3.1 in]{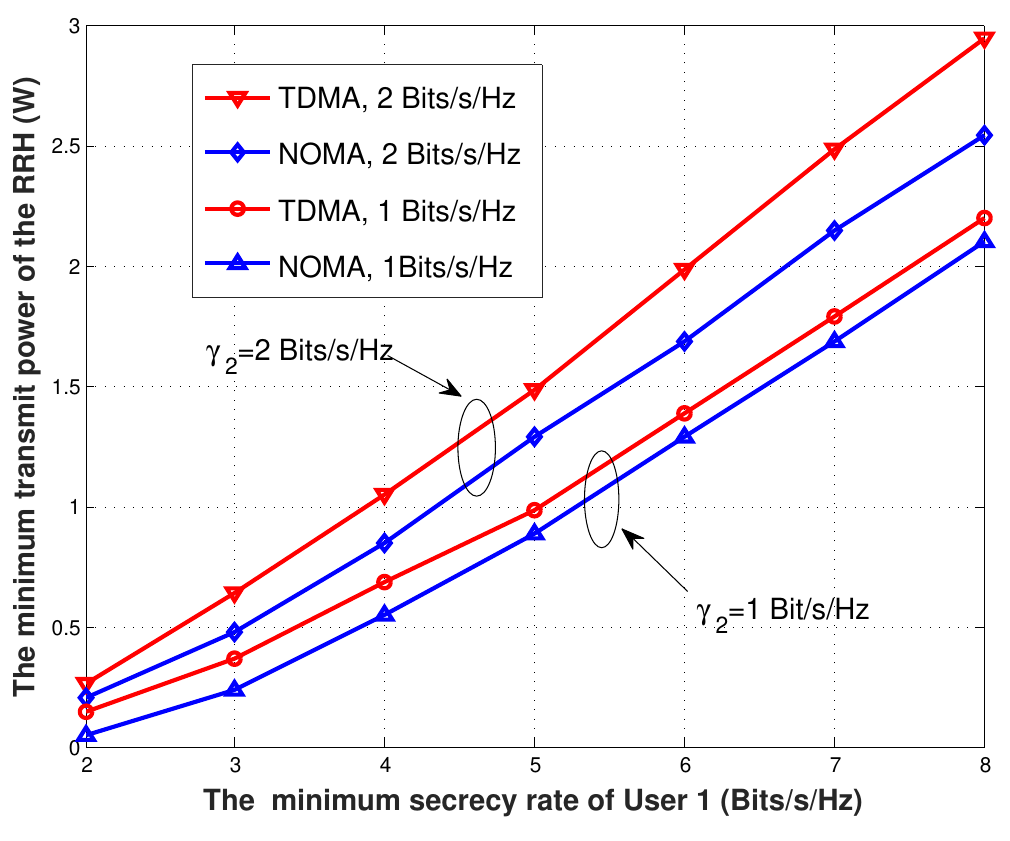}
\put(-310,-10){\footnotesize{(a)}}
\put(-100,-10){\footnotesize{(b)}}
\caption{ (a) Operation modes and wireless charging receiver structure. $T$ denotes the frame duration; $\eta$ represents the time switching factor; $\alpha$ denotes the power splitting factor; $N$ denotes the total number of antennas of the receiver and $M$ represents the number of energy harvesting antennas; (b) The minimum transmission power of the micro RRH versus the minimum secrecy rate of User 1 achieved by using NOMA and TDMA.} \label{fig1}
\end{figure}

As shown in Fig. 5(a), there are two operation modes for RRHs when RRHs are identified as the energy sources for NOMA UEs, namely, the wireless charging mode and the simultaneous wireless charging and information transmission mode. In the first mode, RRHs only provide energy supply for NOMA UEs in the downlink and forward information from NOMA UEs to the BBU pool in the uplink. In the second mode, RRHs transfer energy and transmit information to NOMA UEs simultaneously in the downlink, and harvest energy from NOMA UEs and forward information from NOMA UEs to the BBU pool simultaneously in the uplink. The structures of RRHs and UEs are relatively simple in the first mode while a higher hardware implementation complexity at RRHs and UEs is required in the second mode.

For the second mode, in order to practically realize simultaneous wireless charging and transmission, the received signal has to be split into two parts, e.g., one for energy harvesting (EH) and one for information decoding (ID). Depending on the splitting domain (time, power, antenna), there are different protocols for achieving signal splitting. For the time-domain protocol, the RRH receivers and UE receivers switch in time between EH and ID. For the power-domain protocol, the received signals are split into two signal streams with different power levels by equipping with a power splitting component in the receivers, e.g., one stream for EH and one for ID. Finally, the antenna domain protocol requires RRH receivers or UEs receivers to equip antenna array, and the antenna array is divided into two groups, one group for realizing EH and the other group for achieving ID. The selection of the signal splitting protocol depends on the affordable hardware implementation ability and the scale requirement in the receiver. For example, if the receiver has a low hardware implementation ability and requires a small size (e.g, wireless sensor), the time-domain protocol is deemed more appropriate.

Fig. 5(b) shows the minimum transmission power of the micro RRH versus the minimum secrecy rate of UE 1 achieved by using NOMA and TDMA. The simulation results consider a micro NOMA system with energy harvesting. The micro RRH provides simultaneous wireless information and power transfer (SWIPT) service for two NOMA users and an energy harvesting receiver (EHR). Due to the broadcasting nature and the dual function of radio frequency signals, the combination of wireless charging with NOMA is susceptible to eavesdropping. In the simulation, physical layer secure techniques are applied to achieve secure communications. The minimum secrecy rate requirement of UE 2, $\gamma_2$,  is set as 1 Bits/s/Hz and 2 Bits/s/Hz. The minimum harvesting energy requirement of the EHR is set to be 10 dBm. As shown in Fig. 5(b), the minimum transmission power of the micro RRH achieved by using NOMA is lower than that required by using TDMA. This indicates that NOMA is more energy-efficient than TDMA.
\subsection{Cooperative Transmission with NOMA}
Cooperative relay transmission has been recognized as an effective technique to extend the service coverage, expand  the system capacity, and improve EE by exploiting spatial diversity, especially when the transmit power and size of devices are limited. Cooperative relay transmission with NOMA in H-CRANs  allows UEs with weak channel condition to consider UEs with strong channel condition as relays. In the downlink, RRHs transmit information to UEs, and UEs with strong channel conditions can help RRHs to relay information to UEs with weak channel conditions. In the uplink, UEs with strong channel conditions can be identified as relay to help UEs with weak channel conditions transmit information to RRHs. This combination is more promising in H-CRANs since inter-cell and intra-cell interference can be effectively controlled even when with massive connectivities. Moreover, the linear signal processing and relay selection can be performed in the BBU pool, which can simplify the structure of RRHs and improve EE at RRHs. Additionally, this combination may play a very prominent role in H-CRANs where mmWave communications are exploited since signal blockage can be reduced by decreasing the transmission distance.

The  implementation complexity and the communication overhead of applying  cooperative relay transmission with NOMA in H-CRANs increase with the number of NOMA UEs and the relay RRHs. It can be very  challenging when there are massive NOMA UEs and relay RRHs. One promising solution is to select and pair nearby NOMA UEs and RRHs to realize cooperative transmission. The pairing procedure can be performed in either a centralized or a distributed manner. For the centralized pairing manner, the BBU pool collects all the required CSI and location information from all UEs and RRHs, and then group them into clusters. For the distributed manner, RRHs group nearbyb UEs as a cluster based on the CSIs and location information from UEs. Generally speaking, a higher performance gain can be obtained by using the centralized pairing manner at the cost of a high overhead, whereas the distributed pairing manner has advantages in the flexibility and mobility but with a lower performance gain.
\subsection{D2D Communications with NOMA}
With the ever increasing popularity of IoT applications such as smart grid networks, intelligent homes and integrated transportation systems, future 5G systems will encounter enormous direct communications among devices. Different from the conventional cellular network where the communication among devices must go through a base station even when the distances among devices are very short, D2D communications enable devices to directly connect with each other. It can extensively improve SE, EE and extend the battery lifetime of devices. In H-CRANs,  D2D with NOMA can enable  direct communication among multiple UEs \cite{Z. Zhang}. For example, as shown in Fig. 1, UE$_{1}$ can use the combination of NOMA with D2D communications to connect with UE$_{2}$ and UE$_{3}$. The direct connection between UE$_{1}$ and UE$_{3}$ cannot achieve due to the poor channel condition, but UE$_{1}$ can connect with UE$_{3}$ with the help of UE$_{2}$. In this case, UE$_{2}$  with a strong channel condition is assumed to have full-duplex D2D communications.

There are two ways to realize D2D communications with NOMA in H-CRANs, namely, in-band mode and out-band mode.  The in-band mode enables D2D UEs to share the same spectrum with H-CRANs links, whereas D2D UEs exploit the unlicensed spectrum for the out-band mode. The motivation of using the in-band mode is that the interference of the unlicensed spectrum is uncontrollable and the interference caused to H-CRANs links is tolerable. For the out-band mode, mmWave spectrum bands can be favorable candidates. These two modes have their respective pros and cons, and neither is  deemed absolutely better than the other one. The challenge of achieving D2D with NOMA in the in-band mode is the interference management and thus an adaptive power allocation strategy is of utmost importance. For the out-band mode, the impediment  may come from the severe shadowing and blockage, which are quite normal in mmWave spectrum bands. Another challenge for the two modes is the requirement of massive device accesses, leading  to an intolerable overhead.
\section{Challenges and Open issues}
To enable  the above mentioned technologies  in reality, there are many challenges and issues that  need to be addressed. Some of these challenges and open issues are discussed in the following.

\textbf{Massive MIMO \& NOMA:} Due to the usage of non-orthogonal resources, the pilot contamination  in massive MIMO may be severe. Although a blind estimation algorithm is efficient to address this problem in OMA systems, it has not been explored in NOMA systems with massive MIMO. To implement the combination of NOMA with massive MIMO in practice, it is  an interesting open issue to study how to design a blind estimation algorithm to  overcome the pilot contamination problem. Moreover, the design of optimal precoding for suppressing inter-cell interference requires a large number of CSI but it is extremely difficult to obtain perfect CSI due to the overhead and delay constraints. Thus, how to design robust precoding schemes under the practically imperfect CSI is another challenging  problem.

\textbf{CR \& NOMA:} The most challenging problem  for NOMA and CR in H-CRANs is how to control the interference caused by the simultaneous access of multiple unlicensed UEs in order to protect the QoS of the licensed UEs. Specifically, when an opportunistic access mode is used, how to utilize the non-orthogonal characteristic to improve the performance of spectrum sensing needs to be explored. When a spectrum-sharing mode is employed, RRHs simultaneously provide services for multiple unlicensed users at the same band owned by the  licensed users. Efficient resource allocation strategies are needed to  manage the mutual interference among the unlicensed users and licensed users.

\textbf{MmWave communications \& NOMA:} Most of the existing investigations have studied the application of mmWave to point-to-point communications (e.g., the fronthaul link between one RRH and the BBU pool).  How to design novel mechanisms for multiple UEs is an even more interesting yet challenging  research problem. Besides, since the performance gains achieved by using mmW communications highly depend on the directional transmissions and the transmission directions, effective beamforming schemes can be developed  to overcome the vulnerability of shadowing and the intermittent connectivity.

\textbf{Wireless Charging \& NOMA:}  Wireless charging techniques can use the radio frequency (RF) signals as the source for energy harvesting. Thus the transmitted RF signals not only transfer energy for energy harvesting, but also carry data  information. Due to the dual function of RF signals, malicious energy harvesting receivers may exist and intercept the confidential transmitted information. Moreover, due to the multicasting nature of NOMA, malicious UEs with better CSIs can decode the confidential transmitted information for UEs with worse CSIs. Thus, the security of H-CRANs where the combination of wireless charging and NOMA is of great importance. However, there are only a few works that investigate the secure transmission problem for this. How to improve the security of NOMA H-CRANs is a further research issue.

\textbf{Cooperative Transmission \& NOMA:} When  there are multiple candidates for relaying, how to select a best NOMA UE to relay information is an interesting and important  problem. Moreover, when there are massive NOMA UEs and relay RRHs, the design of pairing schemes is critical  and needs to be studied in order to achieve a good tradeoff between the implementation complexity and the transmission capacity.

\textbf{D2D Communications \& NOMA:} D2D UEs can use the same frequency bands of licensed UEs under the inband D2D communications. Thus, it is important to efficiently manage the interference so that D2D communications do not disrupt the services for the licensed UEs, especially when there are massive D2D users. The design of optimal power allocation schemes and user clustering are interesting open issues for the implementation of the combination of D2D Communications with NOMA.
\begin{figure}[!t]
\centering
\includegraphics[height=3.2in,width=6.4in]{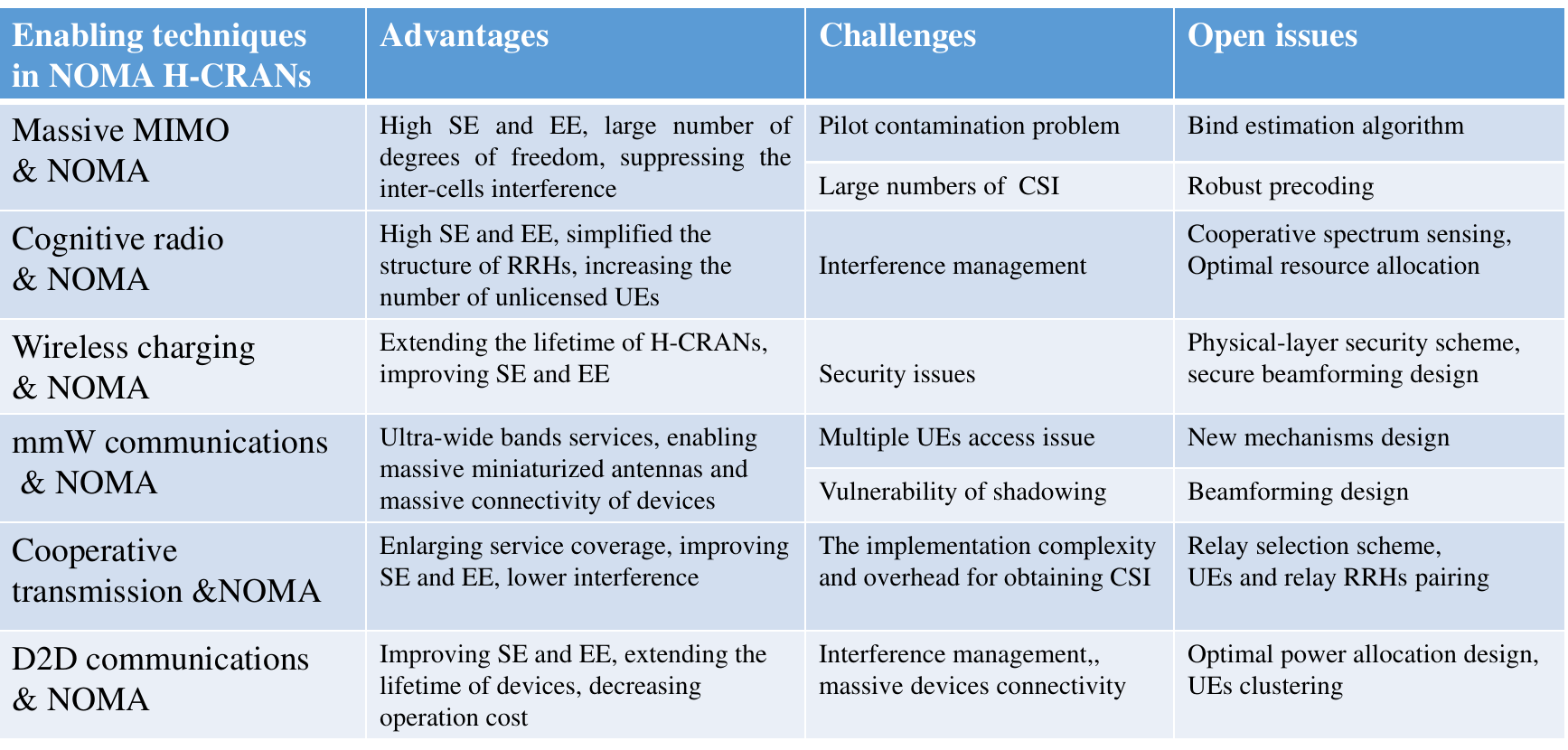}
\caption{Comprehensive survey of enabling techniques for NOMA H-CRANs} \label{fig.1}
\end{figure}

Fig. 6  comprehensively summarizes the advantages, challenges and open issues of each enabling technology.
\section{Conclusions}
The article proposed a new framework for NOMA H-CRANs that aim to provide a high EE, a massive UEs connectivity, and a high system capacity. Several promising enabling technologies for NOMA H-CRANs were presented. These technologies include combining massive MIMO, CR, mmW communications, wireless charging, cooperative transmission, and D2D communications with NOMA in H-CRANs. The implementation challenges and open issues were discussed. The preliminary performance study shows that the proposed NOMA H-CRANs framework and the presented enabling technologies are promising in achieving high EE and SE in the advanced 5G wireless communication systems.

Fuhui Zhou received the Ph. D. degree from Xidian University, Xi¡¯an, China, in 2016.
He joined the School of Information Engineering, Nanchang University, in 2016. He
worked as an international visiting Ph. D student of the University of British
Columbia from 2015 to 2016. Since Aug. 2017, He has worked as a research fellow at Utah State University.
His research interests focus on 5G wireless networks, green communications, cognitive radio, NOMA, energy harvesting, physical layer security.

Yongpeng Wu received the B.S. degree in telecommunication engineering from
Wuhan University, Wuhan, China, in July 2007, the Ph.D. degree in communication and signal processing
with the National Mobile Communications Research Laboratory, Southeast University, Nanjing,
China, in November 2013. He is currently a Tenure-Track Associate
Professor with the Department of Electronic Engineering, Shanghai Jiao Tong University, China. Previously,
he was senior research fellow with Institute for Communications Engineering, Technical University of Munich, Germany
and the Humboldt research fellow and the senior research fellow with Institute for Digital Communications, Universit\"{u}t Erlangen-N$\ddot{u}$rnberg, Germany. During his doctoral studies, he conducted cooperative research at the Department of Electrical Engineering, Missouri University of Science and Technology, USA. His research interests include massive MIMO/MIMO systems, physical layer security, signal processing for wireless communications, and multivariate statistical theory.

Rose Qingyang Hu  is a Professor of Electrical and Computer Engineering Department at Utah State
University. She received her B.S. degree from University of Science and Technology of China,
her M.S. degree from New York University, and her Ph.D. degree from the University of Kansas. She
has more than 10 years of R\&D experience with Nortel, Blackberry and Intel as a technical manager,
a senior wireless system architect, and a senior research scientist, actively participating in industrial
3G/4G technology development, standardization, system level simulation and
performance evaluation. Her current research interests include next-generation
wireless communications, wireless system design and optimization, green
radios, Internet of Things, Cloud computing/fog computing, multimedia
QoS/QoE, wireless system modeling and performance analysis. She has
published over 170 papers in top IEEE journals and conferences and holds
numerous patents in her research areas. Prof. Hu is an IEEE Communications
Society Distinguished Lecturer Class 2015-2018 and the recipient of
Best Paper Awards from IEEE Globecom 2012, IEEE ICC 2015, IEEE
VTC Spring 2016, and IEEE ICC 2016.

Yuhao Wang received the Ph.D. degree in Space Exploration Technology from Wuhan University, Wuhan, China, in 2006. He was a Visiting Professor with the Department of Electrical Communication Engineering, University of Calgary, Calgary, AB, Canada, in 2008, and the China National Mobile Communication Research Laboratory, Southeast University, Nanjing, China, from 2010 to 2011. He is currently a Professor with the Cognition Sensor Network Laboratory, School of Information Engineering, Nanchang University, Nanchang, China. He is an IET Fellow in 2016. He serves for many international journals, such as International Journal of Advanced Robotic Systems. His current research interests include Wideband wireless communication and radar sensing fusion system, channel measurement and modeling, nonlinear signal processing, smart sensor, image and video processing and machine learning.

Kai-Kit Wong received the B.Eng,
M.Phil., and Ph.D. degrees from the Hong Kong University of Science and Technology, Hong Kong,
in 1996, 1998, and 2001, respectively, all in electrical
and electronic engineering. He is currently
a Professor of Wireless Communications with the
Department of Electronic and Electrical Engineering,
University College London, U.K. Prior to this,
he took up faculty and visiting positions at the
University of Hong Kong, Lucent Technologies,
Bell-Labs, Holmdel, NJ, U.S., the Smart Antennas
Research Group of Stanford University, and the Department of Engineering,
the University of Hull, U.K. He is a Fellow of IEEE and IET. His current research interests include Game-theoretic cognitive radio networks, cooperative communications, multiuser communications theory, physical-layer security
Massive MIMO, and energy-harvesting wireless communications.

\end{document}